\documentclass[%
 preprint,
 amsmath,amssymb,
 aps,
prl,
]{revtex4-1}
\usepackage{graphicx}
\usepackage{dcolumn}
\usepackage{bm}

\begin{document}


\title{Probing collective oscillation of $d$--orbital electrons at the nanoscale }

\author{Rohan Dhall}
\affiliation{Department of Materials Science and Engineering, North Carolina State University, Raleigh, North Carolina 27695, USA}
\author{Derek Vigil-Fowler}
\altaffiliation{National Renewable Energy Laboratory, Golden, Colorado 80401, USA}
\author{J. Houston Dycus}%
\affiliation{Department of Materials Science and Engineering, North Carolina State University, Raleigh, North Carolina 27695, USA}
\author{Ronny Kirste}
\altaffiliation{Adroit Materials, Inc., Cary, North Carolina 27518, USA}
\author{Seiji Mita}
\altaffiliation{Adroit Materials, Inc., Cary, North Carolina 27518, USA}
\author{Zlatko Sitar}
\affiliation{Department of Materials Science and Engineering, North Carolina State University, Raleigh, North Carolina 27695, USA}
\author{Ramon Collazo}
\affiliation{Department of Materials Science and Engineering, North Carolina State University, Raleigh, North Carolina 27695, USA}
\author{James M. LeBeau}
 \email{jmlebeau@ncsu.edu}
\affiliation{Department of Materials Science and Engineering, North Carolina State University, Raleigh, North Carolina 27695, USA}




\date{\today}

\begin{abstract}

Here we demonstrate that high energy electrons can be used to explore the collective oscillation of $s$, $p$, and $d$  orbital electrons at the nanometer length scale. Using epitaxial AlGaN/AlN quantum wells as a test system, we observe the emergence of additional features in the loss spectrum with increasing Ga content. A comparison of the observed spectra with ab--initio theory reveals the origin of these spectral features is attributed to 3$d$--electrons contributed by Ga. We find that these modes differ in energy from the valence electron plasmons in Al$_{1-x}$Ga$_x$N due to the different polarizability of the $d$ electrons. Finally, we study the dependence of observed plasmon modes on Ga content, lending insight into plasmon coupling  with electron–-hole excitations.

\end{abstract}

\maketitle



%
%
%

The collective oscillation of free electrons (plasmons) is well described by the Drude model, which is based on the classical equations of motion for a free electron gas \cite{jackson1999classical,pines1966elementary}. In response to an externally applied electromagnetic field, the polarization of the electron gas generates a restoring force, creating longitudinal oscillatory modes in the Fermi sea. These modes can be sustained in any material at a characteristic frequency where the real part of the dielectric function goes to zero \cite{ziman1972principles}, known as the plasma frequency, $\omega_p$. For a free electron gas, $\omega_p = \sqrt{\frac{ne^2}{\epsilon m}}$ , where $n$ is the density of electrons, $e$ and $m$ are the charge and mass of an electron, and $\epsilon$ is the permittivity. In spectroscopic studies of simple metals, this energy is also referred to as the ``ultraviolet transmission limit'', since  metals reflect photons with energy less than $\hbar\omega_p$ and transmit above this energy. This energy is very sensitive to the density of electrons in the Fermi sea undergoing oscillation, spanning a spectral range from the deep UV to far infrared. In particular, as the density of electrons varies from ~10$^{22}$ cm$^{-3}$ (in simple metals) to ~10$^{10}$ cm$^{-3}$ in (in dielectrics), this plasma energy varies from $\sim$4 eV to $\sim$4$ \mu$eV. In a typical semiconducting material such as silicon, the valence band is filled at room temperatures, with each atom contributing four electrons to the Fermi sea. This gives rise to a plasmon of energy $\sim$17eV. Since the electron density in the conduction band of a semiconductor is low at room temperature, collective oscillations of those electrons lie in the infra--red, and can only be probed by sensitive IR spectroscopy.  Despite spanning this large spectroscopic range, the fundamental physics of these excitations is described with considerable success using the simple free electron model.

Within the free electron model, several important assumptions are made: all electrons in the system are treated as identical (or, equally ``free'') , and the interactions between these electrons, as well as the influence of the periodic crystal potential generated by the ``ion cores'' in the solid are ignored \cite{ziman1972principles}. Consequently, any interaction between ``collective modes'' i.e.~plasmons, and the spectrum of single electron states (interband transitions) are not considered in the Drude model. In most metals, ignoring the periodic crystal potential is justified, since typical plasma frequencies (5-50 eV) are significantly higher than the energy of the single particle states (electron--hole pairs), and there is minimal admixing of plasmons and electrons. In wide band gap semiconductors, such as GaN ($E_g$ = 3.4 eV) or AlN ($E_g$ = 6 eV), however, solid--state effects are pronounced.

While plasmons in metals present a well defined, long lived excitation, the situation is considerably more complex in semiconductors where the influence of the crystal potential can not be ignored. The resultant plasmon coupling with the single particle electron--hole pair excitations (Landau damping), is an important decay channel through which plasmons dissipate energy, thus reducing plasmon lifetime. Several applications of plasmons have been proposed, such as the creation of a perfect lens \cite{PhysRevLett.85.3966}, invisibility cloaking \cite{alu2005achieving} or as conduits of on--chip signals which can interface naturally with optics \cite{ozbay2006plasmonics}. The realization of these exciting applications, however, requires knowledge of the  various scattering mechanisms that affect plasmons and their relative importance. By mapping such plasmon modes at the nanometer scale, the role of interfaces, surfaces, and compositional fluctuations can be connected to plasmon scattering mechanisms in inhomogeneous media.



Features associated with plasmon creation are observed in spectroscopic techniques such as X-Ray absorption spectroscopy (XAS) and electron energy loss spectroscopy (EELS). While XAS typically offers better spectral resolution, EELS is capable of providing information at the atomic scale by scanning a finely-focused high energy electron probe across the sample \cite{muller1999electronic}. For plasmon imaging, the spatial resolution is limited to nanometer length scales by delocalization of the plasmon excitation  \cite{egerton1986eels}.  In the case of metal nanoparticles, this technique allows experimentalists to probe the dependence of the plasmon energy on particle shape  \cite{nelayah2007mapping,kociak2000plasmons}, the coupling and energy splitting of plasmonic modes of particles separated by a nanoscale gap  \cite{koh2009electron,agrawal2017resonant}, the mapping of surface plasmons  \cite{bosman2007mapping}, and the quantization of the plasmon modes \cite{scholl2012quantum}. A collection of such results are presented in the review by Colliex et al.~\cite{colliex2016electron}. 


In this Letter, we analyze the plasmon loss signal in AlN/Al$_{1-x}$Ga$_x$N quantum wells using scanning transmission electron microscopy (STEM) and EELS. We show that in addition to the primary plasmon peak (due to collective motion of the valence band electrons), two additional spectral features appear when Ga is present. We study the energies of these modes as a function of Ga composition and find that while the primary valence electron mode downshifts in energy, the additional modes shift to higher energies with increasing Ga fraction. Comparison with ab--initio calculations shows that these additional spectral features arise from the collective oscillation of Ga $d$ orbital electrons. Through this insight, we explain the dependence of plasmon energy on Ga composition, as well as changes in the peak linewidths, which sheds light on the relative importance of different plasmon scattering mechanisms.

AlN/Al$_{1-x}$Ga$_x$N quantum well structures were grown by metal--organic chemical vapor deposition (MOCVD) on vicinal c-plane AlN and sapphire substrates as described previously \cite{dalmau2011growth,bryan2016role}.  The quantum well width was varied between 2 nm and 50 nm and the composition range was varied between AlN (x = 0) and GaN (x = 1). Electron microscopy cross-sectional samples were wedge polished  to electron transparency using an Allied Multiprep system \cite{voyles2003imaging}.  For final thinning, the samples were argon ion milled (Fischione Model 1050) at energies decreasing from 2 keV to 200 eV. A probe--corrected FEI Titan 60--300 kV STEM/TEM equipped with an X--FEG source was used for imaging and spectroscopy. Energy dispersive X-Ray spectroscopy was used to determine the composition of each quantum well, and was found to be in agreement with the expected values from the MOCVD conditions.  The electron beam acceleration voltage was 80 keV to increase the cross--section of plasmon creation. Electron energy loss spectroscopy (EELS) was performed with a Gatan Enfinium ER spectrometer and with an energy spread of approximately 0.9 eV.  As measured with the EELS log-ratio method, the sample thickness varied between 40 and 80 nm \cite{malis1988eels}.  For plasmon energy and linewidth analysis, the spectra were first aligned relative to the zero--loss signal. The effects of plural scattering were then removed using Fourier-log deconvolution \cite{egerton1986eels}.  The plasmon loss peak was fit to the Drude model, where the permittivity of an electron gas with density $n$ is given by: 
\begin{equation}
    \epsilon(\omega) = 1 - \frac{\omega_p^2}{\omega^2 + i\omega/\tau}
\end{equation}
and the loss function is given by Im$\{-1/\epsilon\}$, which corresponds to a Lorentzian peak centered around the plasmon energy, $\hbar\omega_p$, having a linewidth given by $\hbar/\tau$, where $\tau$ is the plasmon lifetime.   

\begin{figure}
\begin{center}
\includegraphics[width = 0.5\textwidth]{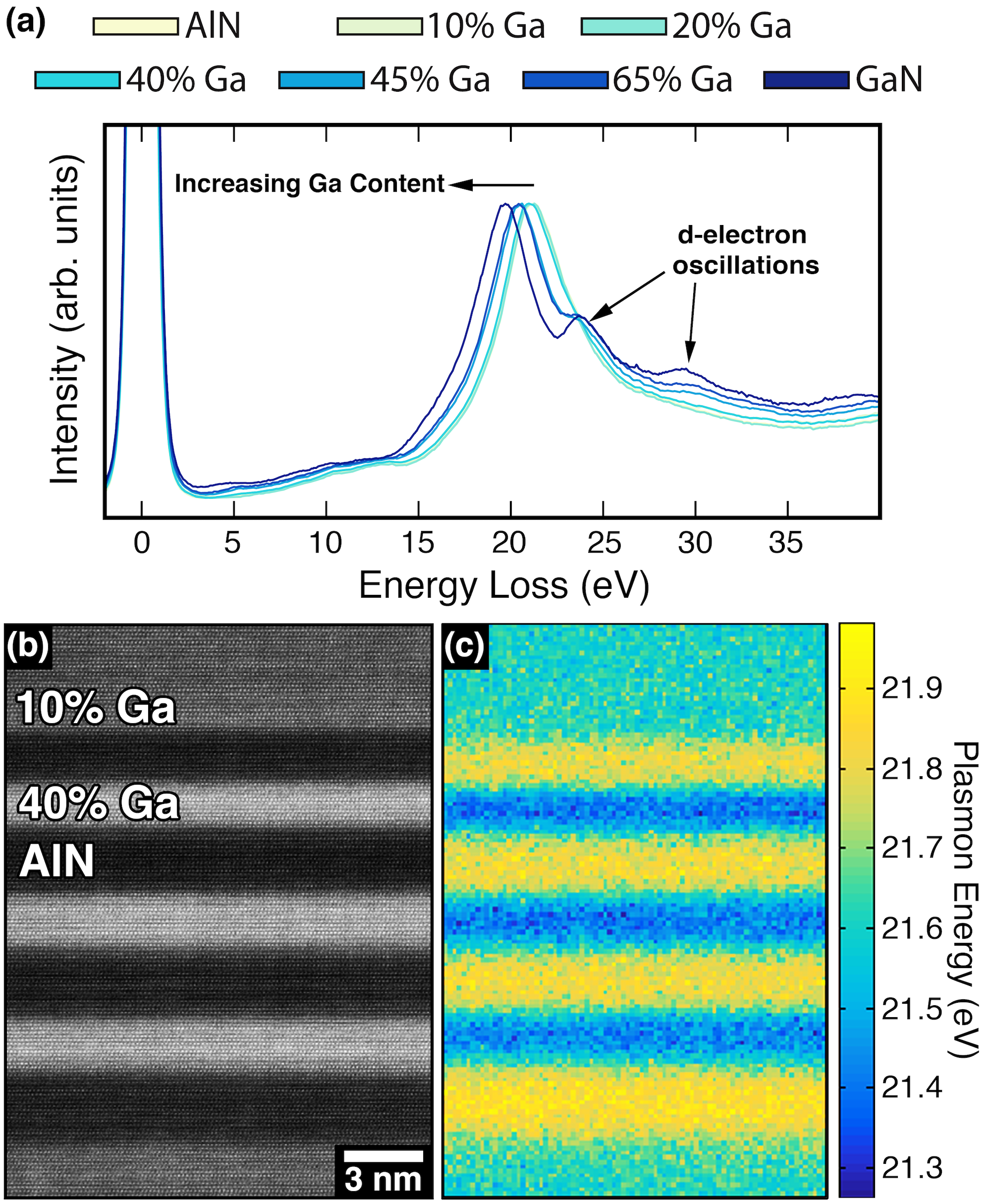}
\caption{ (a) Energy loss spectra at varying compositions of Al$_{1-x}$Ga$_x$N. (b) HAADF STEM overview of the quantum well structures used to obtain the spectra in (a). (c) The corresponding plasmon energy shift across the region in (b) revealing a downshift in the plasmon energy as Ga fraction increases. }
\label{fig:haadf_spec}

\end{center}
\end{figure}

The EEL spectra shown in Figure~\ref{fig:haadf_spec}(a) are acquired from quantum well structures with various Al$_{1-x}$Ga$_x$N compositions. The different quantum wells, as in Figure \ref{fig:haadf_spec}(b), are otherwise similar in terms of sample thickness.  The image intensity variation in Figure~\ref{fig:haadf_spec}(b) results from the atomic number sensitivity of HAADF STEM and scales with Ga content. The variation of the valence plasmon energy across this quantum well structure is shown in  Figure~\ref{fig:haadf_spec}(c).  Two trends are established from this data. First, increasing Ga incorporation leads to a downshift of the ``primary'' plasmon peak (p$_1$) to lower energy. Second, additional energy loss signatures emerge at $\sim$24 (p$_2$) and $\sim$28 (p$_3$) eV. The downshift in the energy of the primary plasmon is a consequence of the decreasing band gap and the increasing unit cell volume with Ga content. Since both Al and Ga are group III elements, they both contribute 3 valence electrons per atom to the crystal. Hence, this increase in unit cell size leads to a decrease in valence electron density and consequently, a downshift in the plasma frequency (given by $\omega_P = \sqrt{\frac{ne^2}{\epsilon_0 m_e}}$). By mapping the position, width, and amplitude of these plasmon modes across the quantum well structures, one can spatially map the Ga fraction of these Al$_{1-x}$Ga$_x$N alloys. This offers a  complementary technique to methods such as EDS or quantitative STEM imaging for identifying the alloy composition \cite{eljarrat2013insight,amari2011nanoscale} .


\begin{figure*}
\centering
\includegraphics[width=.9 \textwidth]{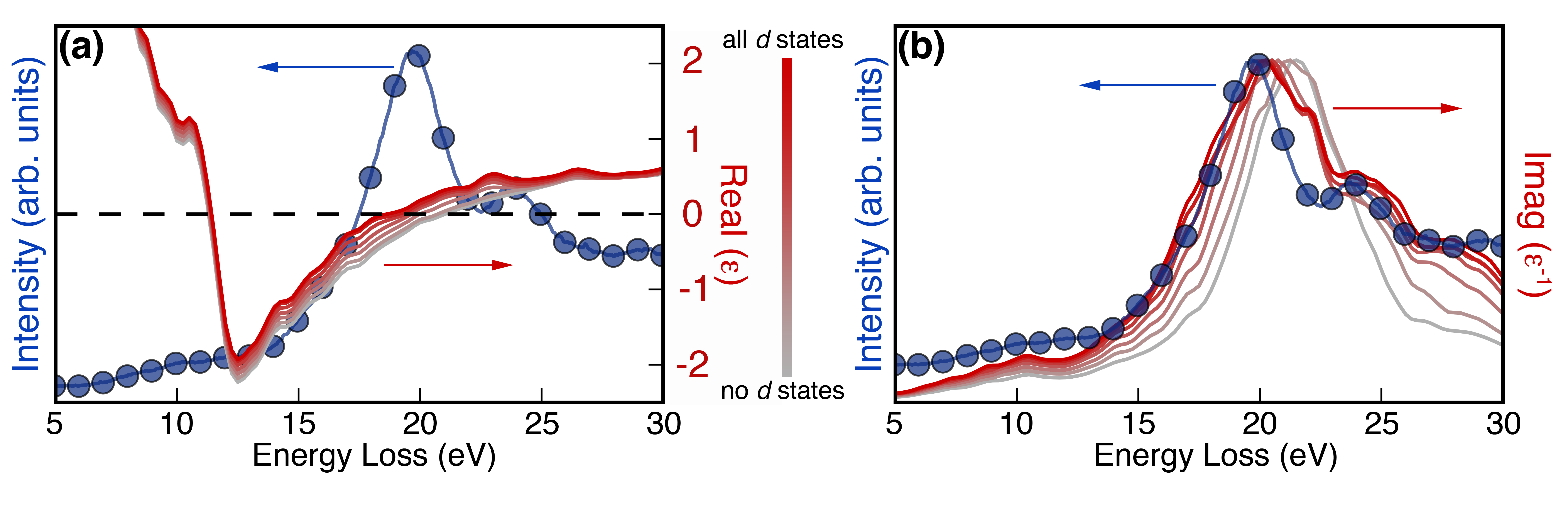}
\caption{Evolution of the (a) calculated  real component of the dielectric function  and (b) loss function of GaN with varying number of Ga $d$ states taken into account. The experimentally recorded EEL spectrum of GaN is also shown. }
\label{fig:GWtheory}
\end{figure*}


The emergence of two additional excitations at 24 eV and 28 eV can be understood by considering the difference in orbital configurations of Ga and Al. Ga contributes $d$ electrons to the Fermi sea due to its occupied 3--$d$ shell, while Al does not. To explore this, we examine results from ab--initio calculations of the Al$_{1-x}$Ga$_x$N dielectric response in the random phase approximation. Many body effects are taken into account using the GW approximation. Rather than Drude's classical description of a ``free--electron'' gas, we apply the Lindhard dielectric response theory \cite{ziman1972principles} to provide a rigorous framework for exploring plasmons in semiconducting materials. The Lindhard model describes a many--electron system in the influence of an external perturbation. This perturbation modifies the electron wavefunctions in this system, leading to a redistribution of charges, and hence the creation of an electrostatic potential. By ensuring self consistency between this generated response field and the original perturbation, we arrive at the following expression for permittivity:

\begin{equation}
\epsilon(q,\omega) = 1 +\frac{4\pi e^2}{q^2} \sum_{\bf{k}}\left[\frac{f(\bf{k}) - f(\bf{k}+\bf{q})}{\varepsilon(\bf{k}+\bf{q}) - \varepsilon(\bf{k}) - \hbar\omega + i\hbar\alpha} \right]    
\end{equation}

\noindent where $f(\bf{k})$ represents the probability of occupation of an electronic state with momentum $\bf{k}$ and energy $\varepsilon(\bf{k})$ while $\omega$ is the frequency of the original perturbation, $\bf{q}$ is the scattering wavevector, and $\alpha$ is introduced to avoid singularities in the summation. It should be noted that this summation extends over all electronic states (occupied as well as unoccupied), denoted by their momentum quantum number $\bf{k}$. At energies much higher than the Fermi energy, states do not contribute to the summation as the numerator goes to zero. Using the random phase approximation, we calculate the permittivity of GaN, systematically increasing the number of electronic bands included in the summation \cite{suppdoc}. 

The results of the Lindhard model are shown in Figures~\ref{fig:GWtheory}(a) and (b), where the calculated real part of the dielectric function and the loss function (i.e, $-1/\epsilon(\omega)$ are provided along with experimental data. As the electronic states of the Ga 3--$d$ shell are included in the calculation, we observe the emergence of additional loss signatures at approximately 24 eV and 28 eV, in excellent agreement with experiment. When these states are excluded from the calculation, the loss function of GaN very closely resembles that of AlN. This leads to the conclusion that these additional features arise from the polarization of $d$ orbital electrons contributed to the crystal as Ga in incorporated in the lattice. In fact, upon revisiting the assumptions within the Drude model, one can see this is to be expected. The model Fermi sea used in the Drude picture assumes electrons live in plane waves states, and all electrons exist in the same parabolic band with the same effective mass. When various electronic bands, with different effective masses contribute to the Fermi sea, the emergence of additional modes is to be expected. This is also true for the AlGaN system, where the dispersions and effective masses for $d$ states is very different from those of the $s$ and $p$ orbital states (Figure S1). Similar observations have been made in graphitic systems for instance, where the $\sigma$ and $\pi$ electrons are known to support different plasmon modes \cite{luo2013plasmons}. 

\begin{figure*}
\centering
\includegraphics[width=1\textwidth]{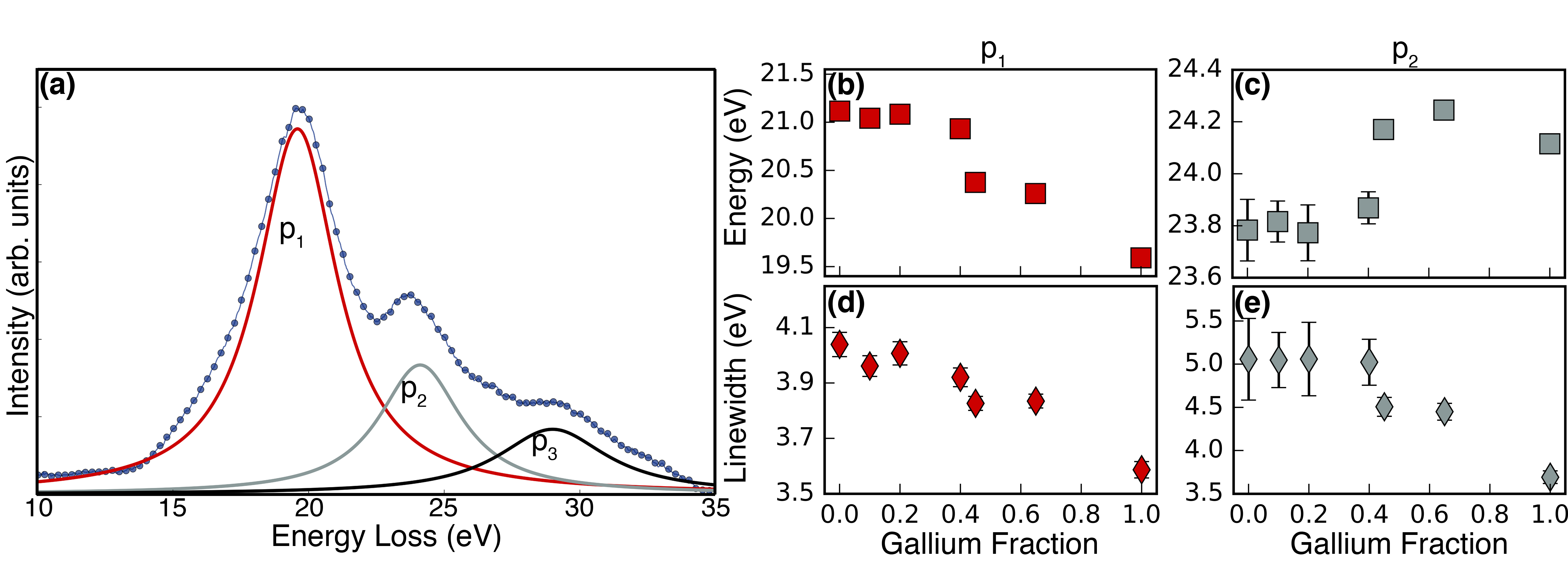}
\caption{(a) Energy loss spectrum for GaN with the three plasmon modes. Each mode is fit to a Lorentizan, which is used to extract (b,c) plasmon energies  and (d,e) linewidths. The fits from p$_3$ suffer from fitting uncertainty and are hence shown only in Figure S2.}
\label{fig:fits}
\end{figure*}



Several different Al$_{1-x}$Ga$_x$N compositions are also used to study the evolution of the energy-loss features as a function of Ga content, as in Figure~\ref{fig:fits}. Explained above, the main bulk plasmon plasmon (p$_1$), originating from the \emph{s} and \emph{p} valence electrons, downshifts as Ga composition is increased. In contrast, both p$_2$ and p$_3$ plasmon modes associated with $d$--electrons, upshift as Ga is added. This provides additional evidence that the origin of these modes is the oscillation of $d$ orbital electrons. As Ga atoms are incorporated into the crystal structure, the number of available $d$ orbital electrons contributing to these modes increases, thus upshifting the plasmon energies of p$_2$ and p$_3$. Since Ga and Al contain the same number of outer shell \emph{s} and \emph{p} electrons, adding Ga does not change the number of electrons contributing to p$_1$, and hence, this peak downshifts with the increasing unit cell size and reduced band gap \cite{wang1996valence}. Note that while the difference in trends of p$_1$ and p$_2$ is obvious, the fitting uncertainty is considerably larger for p$_3$, especially at low Ga compositions. Hence, we have left the discussion of fits to p$_3$ in the supplemental document (Figure S2).   

We also note the trends for the linewidth of these plasmons. As Ga fraction increases, we observe a sharpening of the plasmon peaks p$_1$ and p$_2$.  In principle, these linewidths yield information about the plasmon excitation lifetime \cite{wang1996valence}. Due to the broad energy spread of the electron beam, however, these linewidths provide only a qualitative measure of the various scattering rates. The excitation lifetime is determined by various scattering mechanisms such as interaction with phonons, or decay into electron--hole pairs, as well as disorder scattering due to alloying. If disorder related scattering due to compositional fluctuation were the dominant scattering mechanism, one expects the highest scattering rates at compositions around Al$_{0.5}$Ga$_{0.5}$N. This is not supported by trends presented in Figure~\ref{fig:fits}. The results are, however, consistent with what can be expected if electron--plasmon coupling is dominant. Typically, the scattering rate between two quantum states is of the form $\frac{1}{\tau} = \frac{|V_{AB}|}{\varepsilon_A - \varepsilon_B}$, where $|V_{AB}|$ is the coupling matrix element between states A and B, and $\varepsilon_A$ and $\varepsilon_B$ are the energies of these states. Hence, this scattering rate is inversely proportional to the difference in energies of the states being coupled. In the case here, $\varepsilon_A$ is the energy of the valence electron plasmon (p$_1$), and $\varepsilon_B$ is the energy of an electron--hole pair. Since the band gap of GaN is much smaller than AlN, it is expected that the higher energy single electron states in AlN are more strongly coupled to plasmons than those in the lower band gap GaN. The plasmon linewidth is thus expected to broaden more for AlN than GaN, in agreement with our observation. Recent theoretical studies have predicted similar phenomena in semiconducting systems including emergence of new collective excitations and broadening due to electron--plasmon coupling in semiconductors \cite{lischner2015satellite, vigil2016dispersion, caruso2015band}. A siimilar trend is observed in the linewidth evolution of the plasmon p$_2$, however, the uncertainty from curve fitting is substantial for plasmons p$_3$ due to poor signal-to-noise when Ga content is low. This uncertainty is also likely the cause of the anomalous linewidth evolution of the plasmon centered at $\sim$28 eV (shown in Figure S2). However, one can not rule out that this plasmon may be coupled differently to the single particle excitations owing to symmetry arguments of the electron--plasmon matrix element \cite{dresselhaus2007group}.

The Nitride materials used in this study offer the ideal platform to observe such effects. The cyrstal growth of these materials is now mature to the point that solid--state phenomena are not washed out by defect scattering. Through the combination of experiment and theory, we have revealed the compositional dependence of 3$d$--orbital plasmons using high energy electrons. By quantifying changes in the plasmon energies and linewidths, the influence of solid--state effects, such as electron--plasmon coupling is  revealed in these wide band gap semiconductors. These results further demonstrate the applicability of low-loss EELS to probe subtle solid--state effects.  Coupled with simultaneous atomic resolution STEM imaging, the approach can lead to spatial mapping of electron scattering pathways.


\begin{acknowledgments}
The authors gratefully acknowledge support for this research from the Air Force Office of Scientific Research (FA9550-14-1-0182). JHD acknowledges support by the National Science Foundation Graduate Research Fellowship (DGE-1252376). This work was performed in part at the Analytical Instrumentation Facility (AIF) at North Carolina State University, which is supported by the State of North Carolina and the National Science Foundation (ECCS-1542015). The AIF is a member of the North Carolina Research Triangle Nanotechnology Network (RTNN), a site in the National Nanotechnology Coordinated Infrastructure (NNCI).  

\end{acknowledgments}

\bibliography{bibl.bib}

\end{document}


\title{Supplementary Material for\\ ``Probing collective oscillation of $d$--orbital electrons at the nanoscale''}

\author{Rohan Dhall}
\affiliation{Department of Materials Science and Engineering, North Carolina State University, Raleigh, North Carolina 27695, USA}
\author{Derek Vigil-Fowler}
\altaffiliation{National Renewable Energy Laboratory, Golden, Colorado 80401, USA}
\author{J. Houston Dycus}%
\affiliation{Department of Materials Science and Engineering, North Carolina State University, Raleigh, North Carolina 27695, USA}
\author{Ronny Kirste}
\altaffiliation{Adroit Materials, Inc., Cary, North Carolina 27518, USA}
\author{Seiji Mita}
\altaffiliation{Adroit Materials, Inc., Cary, North Carolina 27518, USA}
\author{Zlatko Sitar}
\affiliation{Department of Materials Science and Engineering, North Carolina State University, Raleigh, North Carolina 27695, USA}
\author{Ramon Collazo}
\affiliation{Department of Materials Science and Engineering, North Carolina State University, Raleigh, North Carolina 27695, USA}
\author{James M. LeBeau}
 \email{jmlebeau@ncsu.edu}
\affiliation{Department of Materials Science and Engineering, North Carolina State University, Raleigh, North Carolina 27695, USA}

\date{\today}
\maketitle

\section{Ab--initio calculations of dielectric response}

To obtain simulated EELS spectra, the inverse dielectric matrix was
calculated using the BerkeleyGW package \cite{deslippe2012berkeleygw}.
The input DFT calculations were done with the Quantum ESPRESSO \cite{giannozzi2009quantum} package
for the mean-field PBE
calculations \cite{perdew1997generalized}. The pseudopotentials used are from the SG15 library \cite{hamann2013optimized}.
The wavefunction cutoff, screened
Coulomb cutoff, and number of bands in the perturbation theory sums were
60 rydberg, 10
rydberg, and 150, respectively. A 12x12x8 sampling of the Brillouin zone
was used.

\begin{figure}[h!]
    \centering
    \includegraphics{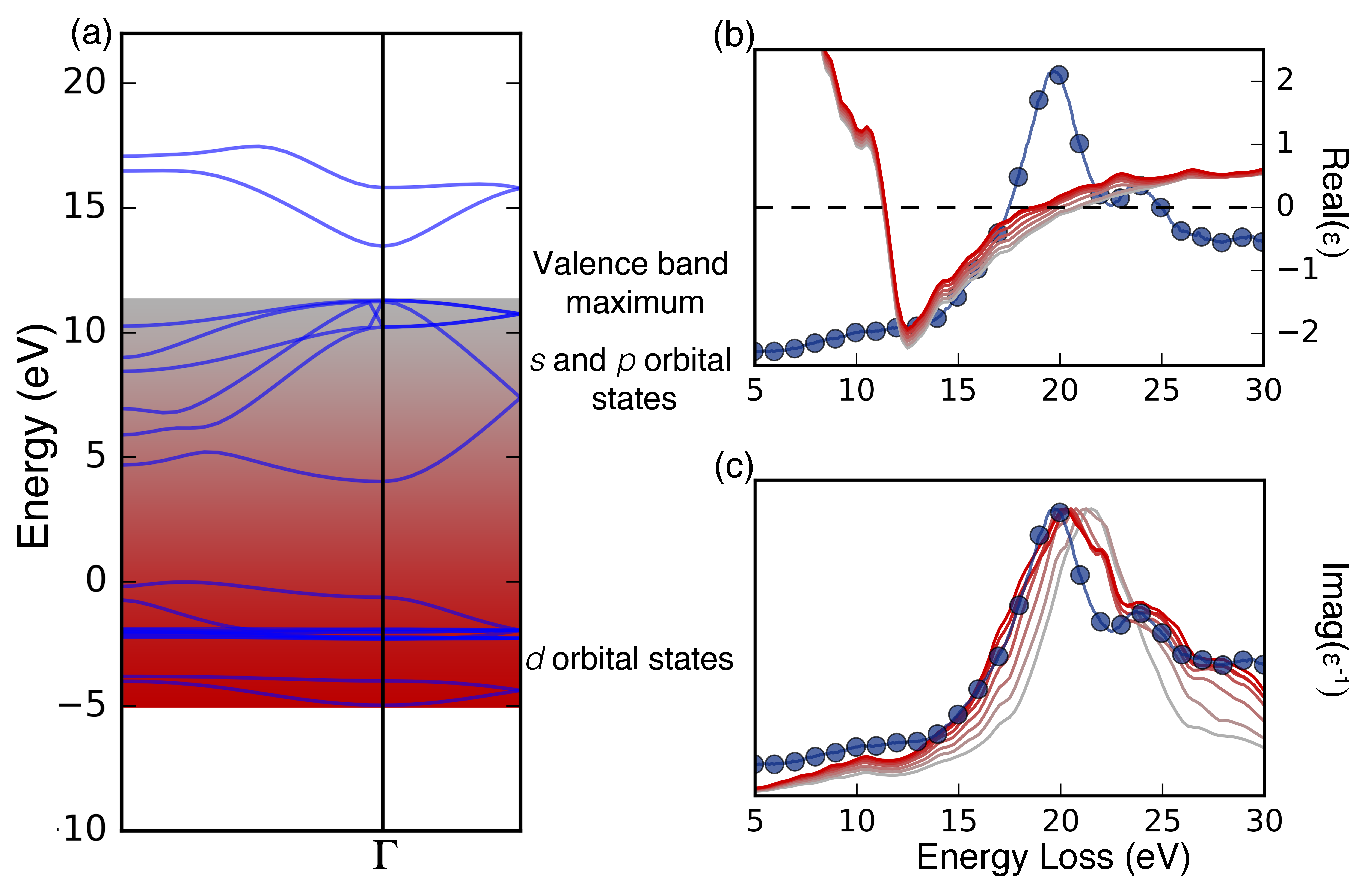}
    \caption{(a) Shows the calculated band structure of GaN. (b) and (c) show the calculated dielectric response function, with increasing contribution from the $d$ states of gallium taken into account.}
    \label{fig:supfig}
\end{figure}

The bandstructure of GaN is shown in Figure~\ref{fig:supfig}(a), where the valence band maximum is near the $\Gamma$ point at ~11 eV. The bands between 5-10eV arise mainly from the $s$ and $p$ states of gallium whereas the $d$ electrons give rise to the deeper lying bands between 0 and -5 eV.    

Figure \ref{fig:supfig} shows the calculated bandstructure in GaN, along with the various bands included in the calculation of the dielectric matrix.The ``full" calculation of the dielectric function includes all the bands, from the valence band maximum, down to the bands which originate from the $d$ states of gallium. By systematically, eliminating specific bands from this calculation, we arrive at the dielectric response of only the $s$ and $p$ orbitals in GaN. The corresponding spectrum shows an absence of additional modes, $p_2$ and $p_3$, which are present in the complete calculation, as evident in Figure~\ref{fig:supfig}(b) and (c) where the real part and inverse of the imaginary part of the dielectric function are shown . 

\begin{figure*}
\centering
\includegraphics[width=1\textwidth]{images/fits5.png}
\caption{(a) Energy loss spectrum for GaN with the three plasmon modes. Each mode is fit to a Lorentizan, which is used to extract (b-d) plasmon energies  and (e-g) linewidths. As is evident, fits from p$_3$ do suffer from significantly greater fitting uncertainty.}
\label{fig:fits}
\end{figure*}


\bibliography{refs.bib}